\documentclass[12pt]{iopart}

\usepackage{graphicx}% Include figure files
\usepackage{iopams}
\usepackage{color}
\usepackage{arydshln}

\newtheorem{theorem}{{\bf Theorem}}

\newtheorem{remark}{{\bf Remark}}

\newtheorem{definition}{{\bf Definition}}

\newcommand{\qed}{\nobreak \ifvmode \relax \else
      \ifdim\lastskip<1.5em \hskip-\lastskip
      \hskip1.5em plus0em minus0.5em \fi \nobreak
      \vrule height0.75em width0.5em depth0.25em\fi}
      
\eqnobysec
\begin{document}
\title[A hierarchy of noncanonical Hamiltonian systems]
{A hierarchy of noncanonical Hamiltonian systems:
circulation laws in an extended phase space}
% \title[Hierarchy of Noncanonical Poisson Manifolds]
% {Hierarchy of Noncanonical Poisson Manifolds and
% Singular Perturbation Unfreezing Casimir Invariants}
%
\author{Z Yoshida$^1$ and P J Morrison$^2$
}

\address{
$^1$Graduate School of Frontier Sciences, University of Tokyo, Kashiwa, Chiba 277-8561, Japan 
\\
$^2$Department of Physics and Institute for Fusion Studies, University of Texas at Austin, Austin, Texas, 78712 USA}
\ead{yoshida@ppl.k.u-tokyo.ac.jp, morrison@physics.utexas.edu}
\date{\today}

\begin{abstract}
The dynamics of an ideal fluid or plasma is constrained by topological invariants 
such as the circulation of (canonical) momentum or, equivalently, the flux of the vorticity or magnetic fields.
In the Hamiltonian formalism, topological invariants restrict the orbits to  submanifolds of the phase space.
While the coadjoint orbits have a natural symplectic structure, the global geometry of the degenerate (constrained) Poisson manifold can be very complex. Some invariants are represented by the center of the Poisson algebra (i.e.,  the Casimir elements such as the helicities), and then, the  global structure of phase space is  delineated by Casimir leaves.
However, a general constraint is not necessarily \emph{integrable}, which precludes the existence of an  appropriate Casimir element; the circulation is an example of such an invariant.
In this work, we formulate a systematic method to embed a Hamiltonian system in an extended phase space; 
we introduce  \emph{mock fields} and extend the Poisson algebra so that the
mock fields are  Lie-dragged by the flow vector.  A mock field defines a new Casimir element, a \emph{cross helicity},
which represents  topological constraints including the circulation.
Unearthing a Casimir element brings about immense advantage in the study of  dynamics and equilibria --- the so-called energy-Casimir method becomes ready  available.  Yet, a mock field does not a priori have a physical meaning.
Here we proffer an interpretation of a Casimir element obtained,  e.g., by such a construction as an \emph{adiabatic invariant} associated with a hidden ``microscopic'' angle variable, and in this way give the mock field  a physical meaning.
We  proceed further and  consider a perturbation of the Hamiltonian by a canonical pair, composed of the Casimir element and the angle,  that causes the topological constraint to be unfrozen.    The theory is applied to the tearing modes of magnetohydrodynamics.

\end{abstract}

\pacs{52.35.We, 45.20.Jj, 47.10.Df, 02.40.Yy}
% 52.35.We Plasma vorticity
% 52.55.Dy General theory and basic studies 
% 45.20.Jj Lagrangian and Hamiltonian mechanics  
% 47.10.Df Hamiltonian formulations 
% 02.40.Yy Geometric mechanics 

% \submitto{\JPA}
\maketitle

%%%%%%%%%%%%%%%%%%%%%%%%%%%%%%%%%%%%%%%%%%%%%%%%%%%%%%%%%%%%%%%%%%%%%%%%%%%%%%%%%%%%%%%%%%%%%%%%
% \newpage
%%%%%%%%%%%%%%%%%%%%%%%%%%%%%%%%%%%%%%%%%%%%%%%%%%%%%%%%%%%%%%%%%%%%%%%%%%%%%%%%%%%%%%%%%%%%%%%%

\section{Introduction}
\label{sec:introduction}

The theory of dynamics can be viewed as built from two elements:  matter and space;
the former is physically an energy, while the latter is mathematically a geometry.
 Hamiltonian mechanics formulates an energy as a Hamiltonian 
that is a function on a phase space $X$,  and 
the geometry of the phase space is dictated by a Poisson bracket
$[F, G ]$ ($F$ and $G$ are functions on $X$), and 
is called a Poisson manifold.
The most basic form of a Poisson manifold is realized by  symplectic geometry, in which case 
 the Hamiltonian mechanics is said to be \emph{canonical}.
A general Poisson manifold, however, may be far more complex than a symplectic manifold, 
and orbits may be constrained by complicated topological invariants
that foliate the phase space into submanifolds (leaves).
Locally, a submanifold can be regarded as a symplectic leaf (Lie-Darboux theorem); 
however, a Poisson operator may have singularities at which leaves bifurcate or intersect.

A nontrivial (non-constant) member $C$ of the \emph{center} of the Poisson algebra
(i.e. $[F,C]=0$ for every $F$) is called a \emph{Casimir element},
which is a constant of motion 
($\rmd C/\rmd t = [H,C] = 0$) for every Hamiltonian $H$.  Contrary to usual 
constants of motion that pertain to symmetries  of a specific Hamiltonian, there are \emph{topological constraints} 
that are  independent of the choice of a Hamiltonian and are  due to the Poisson bracket alone. 
Among various topological constraints, Casimir elements have special importance. 
We call the level-set of a Casimir element a \emph{Casimir leaf},
on which equilibrium points or statistical equilibrium distributions may have 
interesting bifurcated structures, even when the Hamiltonian is simple.
Since the transformation of a Hamiltonian $H$ to an 
\emph{energy-Casimir function} $H_\mu=H-\mu C$ does not change the
dynamics ($\rmd F/\rmd t = [F,H] = [F,H_\mu]$)\,
\cite{Kruskal1958,Hazeltine1984,Morrison1998,Arnold-Khesin},
the equilibrium points (the critical points of $H_\mu$) may bifurcate
when we change $\mu$ as a parameter (or, when we seek equilibria on different Casimir leaves)\,\cite{YD2012}.
Similarly, the Gibbs distribution on a Casimir leaf is given by 
$e^{\beta (H - \mu C)}$,
which can be regarded as a grand-canonical distribution function
($\mu$ is a chemical potential)\,\cite{Y-RT1-2013}.
We note that the equilibrium or the Gibbs distribution function of
a canonical Hamiltonian system can be nontrivial only when the Hamiltonian is a bumpy function,
but this  is not the case for a weakly coupled system like a usual fluid or a plasma.

In the context of the present study, we highlight another distinction of 
Casimir elements among topological constraints. 
In \cite{YM2014}, we proffered an interpretation of a Casimir element as an
\emph{adiabatic invariant} associated with a hidden ``microscopic'' angle variable.
Adding the angle variable to the phase space, we `alchemized' 
the Casimir element into an action variable, which together with the angle variable forms a canonical pair.  Then,  
 perturbing the Hamiltonian by the new canonical variables, we  unfroze the Casimir element. 
By this theory, we  extended  the scope of ideal Hamiltonian mechanics to see what happens
when the orbit is allowed to deviate from the leaves of the Poisson manifold. 
A finite dissipation may break the ideal constraints and free the orbit to move 
among different leaves when a  very small dissipation that does not destroy the basic structure of the Poisson manifold is considered (as opposed to large dissipation that  diminishes the ``dimension'' of the dynamics).
Thus,  ideal constraints can be removed, giving rise to some instabilities.

In this work, we formulate a general systematic method for  embedding a Poisson manifold 
into  a higher-dimensional phase space and, in doing so,  express  the topological constraints 
(restricting important instabilities omitted in the ideal model) in terms of Casimir elements of the embedded system.  
This idea is motivated by  early work \cite{MH84} in which it was observed that adding additional variables to a noncanonical Hamiltonian  theory enriched the Casimir structure and made available more general equilibria for the energy-Casimir method.  
The idea was later used explicitly in the Vlasov-Poisson context in \cite{pjm87}, and our method of embedding is a special case of the general theory of extensions  given in \cite{thiffeault2,thiffeault}.    
Specifically, we introduce a \emph{mock field} by which a local topological constraint
(which cannot be elucidated by the original Casimir elements) is represented as a Casimir element, a \emph{cross helicity} pertinent to the mock field (the reader is referred to \cite{Fukumoto2008,Fukumoto2013} for the original idea of
unifying topological invariants as cross helicities).  
Then, the mock field is the target to be perturbed when one wishes to break topological constraints.

We put the theory to the test by analyzing the equations of ideal magnetohydrodynamics (MHD), which was first shown in \cite{MG80}  to have noncanonical Hamiltonian form on an infinite-dimensional phase space of Eulerian variables.
Alfv\'en's law, that the local magnetic flux on every co-moving surface is a topological invariant, prevents  any change in the linkage of magnetic field lines.  Alternatively this law can be viewed as a rephrasing of Kelvin's circulation law with the magnetic field replacing the vorticity.  Therefore, tearing modes, which grow by creating magnetic islands, are forbidden in an ideal plasma\,\cite{FKR63,Furth63,White83}.  Here we show that the magnetic flux on a co-moving surface is 
the cross helicity pertinent to a Lie-dragged \emph{pure-state}\,\cite{YKY2013} mock field.    Hence, upon unfreezing this cross helicity the local (resonant) magnetic flux can give rise to tearing modes\,\cite{YD2012,YM2014}.

%%%%%%%%%%%%%%%%%%%%%%%%%%%%%%%%%%%%%%%%%%%%%%%%%%%%%%%%%%%%%%%%%%%%%%%%%%%%%%%%%%%%%%%%%%%%%%%%%%
%%%%%%%%%%%%%%%%%%%%%%%%%%%%%%%%%%%%%%%%%%%%%%%%%%%%%%%%%%%%%%%%%%%%%%%%%%%%%%%%%%%%%%%%%%%%%%%%%%
%%%%%%%%%%%%%%%%%%%%%%%%%%%%%%%%%%%%%%%%%%%%%%%%%%%%%%%%%%%%%%%%%%%%%%%%%%%%%%%%%%%%%%%%%%%%%%%%%%
%%%%%%%%%%%%%%%%%%%%%%%%%%%%%%%%%%%%%%%%%%%%%%%%%%%%%%%%%%%%%%%%%%%%%%%%%%%%%%%%%%%%%%%%%%%%%%%%%%
%%%%%%%%%%%%%%%%%%%%%%%%%%%%%%%%%%%%%%%%%%%%%%%%%%%%%%%%%%%%%%%%%%%%%%%%%%%%%%%%%%%%%%%%%%%%%%%%%%
\section{A hierarchy of noncanonical Hamiltonian systems}
\label{sec:hierarchy}

\subsection{Noncanonical Hamiltonian systems and degenerate Poisson manifolds}
\label{sec:Hamiltonian_formalism}

A general Hamiltonian system may be written as
\begin{equation}
\frac{\rmd}{\rmd t} u = \mathcal{J}(u) \partial_{u} H(u),
\label{Hamilton_eq_2}
\end{equation}
where $u$ is the state vector, a member of the phase space $X$ (here a function space), 
$H(u)$ is the Hamiltonian (here a real-valued functional on $X$), and
$\mathcal{J}$ is the Poisson operator (or cosymplectic bivector).
We allow $\mathcal{J}$ to be a function of $u$ on $X$, and write it as $\mathcal{J}(u)$.
We assume that the Poisson bracket, the bilinear product,    
\[
[F,G] =
\langle \partial_u F(u), \mathcal{J} \partial_u G(u)\rangle
\]
is antisymmetric and satisfies the Jacobi identity.

A \emph{canonical} Hamiltonian system is endowed with a \emph{symplectic} Poisson operator where 
\[
\mathcal{J}_c = \left( \begin{array}{cc}
0 & I
\\
-I & 0
\end{array} \right).
\]
However, our interest is in  \emph{noncanonical} systems endowed with
Poisson operators  $\mathcal{J}$ that are inhomogeneous and degenerate
(i.e.,  $\textrm{Ker}(\mathcal{J}(u))$ contains nonzero elements, and its dimension may change depending on the position in $X$).  Since $\mathcal{J}$ is antisymmetric, $\textrm{Ker}(\mathcal{J}(u))=\textrm{Coker}(\mathcal{J}(u))$,
and hence, every orbit is topologically constrained on the orthogonal complement of
$\textrm{Ker}(\mathcal{J}(u))$. 

A functional $C(u)$ such that $[C,G]=0$ for all $G$ is called a Casimir element
(or an element of the center of the Poisson algebra).
If $\mathrm{Ker}(\mathcal{J})=\{0\}$, the case for a canonical Hamiltonian system,  then there is only a trivial element $C=$ constant in the center.
% A Hamiltonian system that has a nontrivial $\mathrm{Ker}(\mathcal{J})$ is said to be \emph{noncanonical}.
% If an element $v\in \mathrm{Ker}(\mathcal{J})$ is ``integrable'' to yield
% a Casimir element $C(u)$ such that $v=\partial_u C(u)$, the
% level-set of $C(u)$, which we call a Casimir leaf, foliates the phase space $X$.
Evidently, a  Casimir element $C(u)$ is a solution to the  differential equation
\begin{equation}
\mathcal{J}(u) \partial_{u} C(u) =0.
\label{Casmir-1}
\end{equation}
When the phase space $X$ has a finite dimension, (\ref{Casmir-1}) is a first-order
partial differential equation.
If $\mathrm{Ker}(\mathcal{J}(u))$ has a constant dimension $\nu$ in an open set $X_\nu\subseteq X$,
we can integrate (\ref{Casmir-1}) in $X_\nu$ to obtain $\nu$ independent solutions,
i.e.,  $\mathrm{Ker}(\mathcal{J}(u))$ is locally spanned by the gradients of 
$\nu$ Casimir elements (Lie-Darboux theorem).
The intersection of all Casimir leaves (the level-sets of Casimir elements)
is the effective phase space, on which $\mathcal{J}(u)$ reduces to a symplectic Poisson operator.

However, the general (global) integrability of (\ref{Casmir-1}) is a mathematical challenge;
the point where the rank of $\mathcal{J}(u)$ changes is a singularity of 
(\ref{Casmir-1})\,\cite{Morrison1998}, 
from which singular (hyper-function) solutions are generated.
Moreover,  because models of a fluids and plasmas are  formulated on an infinite-dimensional phase space, for these systems   (\ref{Casmir-1}) is a functional differential equation.
The reader is referred to \cite{YMD2013} for an example of a singular Casimir element
generated by singularities in a function space.

Our strategy of improving the integrability of (\ref{Casmir-1}) and extending
the set of   topological constraints expressible in terms of Casimir elements 
is to embed the Poisson manifold in higher-dimensional spaces.  
For an element $v\in \mathrm{Ker}(\mathcal{J}(u))$, (\ref{Casmir-1}) demands  a solution in terms of  a \emph{gradient} (exterior derivative) of a scalar potential (0-form) $C(u)$.
Such a solution  is possible only when $v$ is an exact 1-from, or at least  $v$ must be a closed 1-form for the local integrability.  Our  idea is to add extra components to $v$ and make it exact in a higher-dimension space.
Although this description  is a finite-dimensional story,   we will develop an infinite-dimensional theory.
In the next subsection, we see how Casimir elements change as the phase space is extended.

\subsection{Example of two-dimensional vortex dynamics} %-------------------------------------
\label{subsec:2D_example}

\begin{table}
\caption{Hierarchy of two-dimensional vortex  systems.
Here  $\{a,b\}= \partial_y a \partial_x b - \partial_x a \partial_y b$.}
\begin{center}
\begin{tabular}{c|c|c|l}
\hline
system & state vector & Poisson operator  & Casimir elements
\\ \hline
(I)  & $\omega$          & $\{ \omega,\circ\}$ & $~C_0 = \int\rmd^2 x\, f(\omega)$
\\ \hline
(II) & $^t(\omega,\psi)$ & 
$\left( \begin{array}{cc} 
\{ \omega,\circ\} & \{ \psi,\circ\} \\
\{ \psi  ,\circ\} & 0     
\end{array} \right)$ & $\begin{array}{l} 
C_1= \int\rmd^2 x\, \omega g(\psi) \\
C_2= \int\rmd^2 x\, f(\psi) \\
\end{array}$
\\ \hline
(III)& $^t(\omega,\psi,\check{\psi})$ & 
$\left( \begin{array}{ccc} 
\{ \omega,\circ\} & \{ \psi,\circ\} & \{ \check{\psi},\circ\} \\
\{ \psi  ,\circ\} & 0  & 0 \\
\{ \check{\psi}  ,\circ\} & 0  & 0
\end{array} \right)$ & $\begin{array}{l} 
C_2= \int\rmd^2 x\, f(\psi) \\
C_3= \int\rmd^2 x\, h(\psi\check{\psi}) \\
C_4= \int\rmd^2 x\, \check{f}(\check{\psi}) 
\end{array}$
\\ \hline
\end{tabular}
\end{center}
% \vspace*{0.3cm}
\label{table}
\label{table:2D-hierarchy}
\end{table}

In Table\,\ref{table:2D-hierarchy} we compare 
well-known examples of two-dimensional vortex dynamics systems, the Hamiltonian structures of which were given in \cite{aip82,MH84,MM84}.
We denote by $\omega=-\Delta\varphi$ the vorticity with $\Delta$ being  the Laplacian 
and $\varphi\in H^1_0(\Omega)\cap H^2(\Omega)$ for the  two-dimensional Eulerian velocity field  
$\bi{V}=~^t(\partial_y\varphi,-\partial_x\varphi)$.
Given a Hamiltonian
\[
H_{{\rm E}}(\omega) = -\frac{1}{2} \int \rmd^2 x \,\omega \,   \Delta^{-1}\omega ,
\]
the system (I) is the vorticity equation for  Eulerian flow, 
\[
\partial_t \omega + \bi{V}\cdot\nabla\omega =0.
\]
In Table~\ref{table:2D-hierarchy}  we show the Poisson operator and Casimir elements for this system. 

If $\psi$ is the Gauss potential of a magnetic field, i.e., 
$\bi{B}=~^t(\partial_y\psi,-\partial_x\psi)$,
and the Hamiltonian is
\[
H_{{\rm RMHD}}(\omega,\psi) = -\frac{1}{2} \int \rmd^2 x \,\left[\omega  \,  \Delta^{-1}\omega + \psi \, \Delta\psi) \right],
\]
the system (II) is the reduced MHD equation, 
\begin{eqnarray*}
& & \partial_t \omega + \bi{V}\cdot\nabla\omega = \bi{J}\times\bi{B},
\\
& & \partial_t \psi + \bi{V}\cdot\nabla\psi = 0. 
\end{eqnarray*}
In the system (II), $C_0$ is no longer a constant of motion, being replaced by $C_1$ and $C_2$ of Table~\ref{table:2D-hierarchy}.   However, if $\psi$ is a \emph{mock field}, i.e.,  if the Hamiltonian $H$ is independent of $\psi$,
both $\omega$ and $\psi$ obey the same evolution equation; we may assume $\psi=\omega$,
and then, both $C_1$ and $C_2$ reduce into $C_0$.
The constancy of $C_0$ is now due to the symmetry $\partial_\psi H =0$.
To put it in another way, a modification of the Hamiltonian to involve $\psi$
destroys the constancy of $C_0$;
the electromagnetic interaction is a physical example of such a modification.

We can extend the phase space further to obtain a system (III) by adding another field $\check{\psi}$ 
that obeys the same evolution equation as $\psi$.
In the reduced MHD system, $\check{\psi}$ is a mock field, i.e.,  it does not have a direct physical meaning;  however, in the original RMHD context  such a field physically correspond to the pressure in the  high-beta MHD model \cite{MH84} (see also \cite{thiffeault2}).  For this further extended system we obtain the additional Casimir elements $C_3$ and $C_4$ of Table~\ref{table:2D-hierarchy}, as first shown in  \cite{MH84}.

\subsection{Integrability of topological constraints} %-------------------------------------
\label{subsec:integrability}

An interesting consequence of extending the system from (I) to (II) is found in the
\emph{integrability} of the $\textrm{Ker}(\mathcal{J})$, or the topological constraints.
In (I),
\[
\textrm{Ker}(\mathcal{J}(\omega)) = \{ \psi;\, \{\omega,\psi\}=0 \},
\]
which implies that $\psi$ and $\omega$ are related, invoking a certain scalar $\zeta(x,y)$, by
\begin{equation}
\psi = \eta(\zeta), \quad \omega= \xi(\zeta).
\label{general_Kernek}
\end{equation}
As far as $\xi$ is a monotonic function, we may write $\psi = \eta (\xi^{-1}(\omega))$, which
we can integrate to obtain the Casimir element $C_0(\omega)$ with $f(\omega)$ such that 
$f'(\omega) = \eta(\xi^{-1}(\omega))$.
Other elements of $\textrm{Ker}(\mathcal{J}(\omega))$ 
that are given by nonmonotonic $\xi$ are not integrable to define Casimir elements.
Yet, we can integrate such elements as $C_1(\omega,\psi)$ % (or $C_2(\psi)$)
in the extended  space of (II).  In fact, every member of $\textrm{Ker}(\mathcal{J}(\omega))$ can be
represented as $\partial_\omega C_1=g(\psi)$ by choosing $\psi$ in $\textrm{Ker}(\mathcal{J}(\omega))$.
%  such that $\{\omega,\psi\}= 0$.

Similarly, % if $\psi$ of the system (II) is a physical field (like the magnetic flux of the reduced MHD system), 
in the system (II),
we encounter the deficit of the Casimir element $C_2=\int\rmd^2xf(\psi)$
in covering all elements $\,^t(0,\chi)\in \textrm{Ker}(\mathcal{J}(\omega,\psi))$ such that $\{\psi,\chi\}=0$.
By the help of a mock filed $\check{\psi}$, we can integrate every element of 
$\textrm{Ker}(\mathcal{J}(\omega,\psi))$ as $C_3$.

\subsection{Minimum canonization invoking Casimir elements} %-----------------------
\label{subsec:canonization}

If a topological constraint on a noncanonical system is represented by a Casimir element, 
we can define a canonical pair by adding an angle variable; 
then, the Casimir element morphs  into an action variable\, \cite{YM2014}.

Here we consider a finite-dimensional model (which may be regarded as a relevant 
degenerate part of an infinite-dimensional system).
Let $J$ be a Poisson operator (matrix) on an $n$-dimensional 
phase space $X=\mathbb{R}^n$ parameterized by $\bi{z} = (z_1, \cdots, z_n)$.
We assume that $\textrm{Ker}(J)$ has a dimension $\nu$ and
$n-\nu$ is an even number.
We first canonize  $J$ on $X/\textrm{Ker}(J)$.
Let
\[
\bi{z}'= (\zeta_1,\cdots,\zeta_{n-\nu},C_1,\cdots,C_\nu) \in \mathbb{R}^n,
\]
by which $J$ is transformed into a Darboux standard form:
\begin{equation}
J' =
\left( \begin{array}{ccc:c}
~   & ~   &  ~   &  ~   \\
 ~  &J_c  & ~    &  ~   \\
 ~  &   ~ & ~    &  ~   \\
  \hdashline
 ~  &  ~  &  ~   & 0_\nu
\end{array} \right) ,
% \left( \begin{array}{c:ccc}
% 0_\nu & 0  & \cdots &0 \\
% \hdashline
% 0 & J_c & 0  & 0\\
% \vdots & 0   & \ddots & 0 \\
% 0  &  0  & 0  & J_c
% \end{array} \right) ,
\label{J-canonical}
\end{equation}
We can extend $J'$ to an $\tilde{n}\times\tilde{n}$ canonical matrix such that
\begin{equation}
{J}_{ex} =
\left( \begin{array}{ccc:cc}
~   & ~   &  ~   &  ~ & ~  \\
 ~  &J_c  & ~    &  ~ & ~  \\
 ~  &   ~ & ~    &  ~ & ~  \\
  \hdashline
 ~  &  ~  &  ~   & 0_\nu &-I_\nu \\
 ~  &  ~  &  ~   &I_\nu &0_\nu 
\end{array} \right) .
\label{J-canonical-extend}
\end{equation}
% where $I_\nu$ is $\nu\times\nu$ identity.
The corresponding variables are denoted by
\[
{\bi{z}}_{ex} 
% \bi{z}'\oplus\bi{\vartheta} % \bi{C}\oplus\bi{\zeta}
=(\zeta_1,\cdots,\zeta_{n-\nu},C_1,\cdots,C_\nu,\vartheta_1,\cdots,\vartheta_\nu) \in \mathbb{R}^{\tilde{n}} .
\]

This extended Poisson matrix ${J}_{ex}$ is \emph{symplectic}, i.e., 
the extended system is \emph{canonized},
which is in marked contrast to the noncanonical extension discussed in
Sec.\,\ref{subsec:2D_example}.
The noncanonical extension is the first step for  representing  topological constraints by Casimir elements.
Next, we extend the phase space further to canonize the Casimir elements.
By perturbing the Hamiltonian with the added angle variables, we can unfreeze the Casimir elements.
This perturbation brings about an increase in the number of degrees of freedom of the system, and is an example of  a \emph{singular perturbation}.

%%%%%%%%%%%%%%%%%%%%%%%%%%%%%%%%%%%%%%%%%%%%%%%%%%%%%%%%%%%%%%%%%%%%%%%%%%%%%%%%%%%%%%%%%%%%%%%%%%
%%%%%%%%%%%%%%%%%%%%%%%%%%%%%%%%%%%%%%%%%%%%%%%%%%%%%%%%%%%%%%%%%%%%%%%%%%%%%%%%%%%%%%%%%%%%%%%%%%
%%%%%%%%%%%%%%%%%%%%%%%%%%%%%%%%%%%%%%%%%%%%%%%%%%%%%%%%%%%%%%%%%%%%%%%%%%%%%%%%%%%%%%%%%%%%%%%%%%
%%%%%%%%%%%%%%%%%%%%%%%%%%%%%%%%%%%%%%%%%%%%%%%%%%%%%%%%%%%%%%%%%%%%%%%%%%%%%%%%%%%%%%%%%%%%%%%%%%
%%%%%%%%%%%%%%%%%%%%%%%%%%%%%%%%%%%%%%%%%%%%%%%%%%%%%%%%%%%%%%%%%%%%%%%%%%%%%%%%%%%%%%%%%%%%%%%%%%
\section{Topological constraints in ideal magnetohydrodynamics}
\label{sec:MHD}

Hereafter, we consider the example of a noncanonical Hamiltonian system provided by   three-dimensional ideal MHD system. The dynamics is strongly constrained by the magnetic flux conservation on every co-moving surfaces.
Local magnetic fluxes are, however, not always Casimir elements
(in two-dimensional dynamics, some are implied by the Casimir elements $C_2$; see Sec.\,\ref{subsec:2D_example}).
Applying the method of the previous section, 
we extend the system to write local fluxes, which are loop integrals,  as Casimir elements.
In this section, we review the basic formulation, boundary conditions, and the magnetic flux conservation law.

\subsection{Magnetohydrodynamics} %---------------------------------------------------------------
\label{subsec:MHD}

Denoting by ${\rho}$ the mass density, $\bi{V}$ the fluid velocity,
$\bi{B}$ the magnetic field, 
% $m$ the ion mass, %$\Phi= V^2/2 + h$ the total enthalpy,
$h$ the specific enthalpy, the governing equations of magnetohydrodynamics (MHD) are 
\begin{eqnarray}
& & \partial_t{\rho} = -\nabla\cdot(\bi{V} {\rho}),
\label{MHD-mass_conservation}
\\
& & \partial_t \bi{V}=-(\nabla\times\bi{V})\times\bi{V} - \nabla(h+V^2/2)
+{\rho}^{-1}(\nabla\times\bi{B})\times\bi{B},
\label{MHD-momentum}
\\
& & \partial_t\bi{B}=\nabla\times(\bi{V}\times\bi{B}) .
\label{MHD-induction}
\end{eqnarray}
Here we assume a barotropic relation to write the enthalpy $h=h(\rho)$
(which is related to the thermal energy $\mathcal{E}$ by
$h=\partial({\rho}\mathcal{E})/\partial{\rho}$).
The variables are normalized in standard Alfv\'en units with energy densities (thermal $\rho\mathcal{E}$, kinetic $\rho_0V^2$ and magnetic $B^2/2\mu_0$) normalized by a representative magnetic energy density $B_0^2/\mu_0$.
Evidently, the state vector for this system  is ${u}={}^t({\rho},\bi{V},\bi{B})$.

% \subsection{Boundary conditions} %-------------------------------------------------------
We consider a bounded domain $\Omega\subseteq\mathbb{R}^3$ on which the
Hamiltonian (energy) has a finite value.
Here we start with a simply connected $\Omega$
(a multiply connected domain will be discussed in Sec.\,\ref{subsec:cohomology}).
We denote by $\partial\Omega$ the boundary of $\Omega$, which is a smooth two-dimensional
manifold consisting of a finite number of connected components.
% $\Gamma_1,\cdots,\Gamma_m$, i.e. 
% \[
% \partial\Omega=\bigcup_{i=1}^m\Gamma_i.
% \]
Denoting by $\bnu$ the unit normal vector on the boundary $\partial\Omega$,
and by $f|_{\partial\Omega}$ the trace of $f$ onto the boundary $\partial\Omega$,
we assume the following standard boundary conditions
on the flow velocity $\bi{V}$ and the magnetic field $\bi{B}$:
\begin{eqnarray}
\bnu\cdot\bi{V}|_{\partial\Omega} &=& 0,
\label{BC-V} \\
\bnu\cdot\bi{B}|_{\partial\Omega} &=& 0.
\label{BC-B}
\end{eqnarray}
Physically, (\ref{BC-V}) means that the fluid (plasma) is confined in the domain and cannot cross the boundary.
The magnetic field is also confined in the domain; (\ref{BC-B}) is a consequence of 
(in fact, a little more stronger than) a perfectly conducting boundary condition isolating $\Omega$ electromagnetically from the complementary space, which demands that the tangential component of the electric field $\bi{E}$
vanishes on $\partial\Omega$, i.e.
\begin{equation}
\bnu\times\bi{E}|_{\partial\Omega} = 0.
\label{BC-E}
\end{equation}
Writing $\bi{E}=-\partial_t \bi{A} - \nabla\phi$ with a scalar potential $\phi$, we observe, 
for every disk $S\subset\partial\Omega$
(where  $\partial S$ is  the boundary of the disk $S$ and $\btau$ is the
unit tangent vector along $\partial S$),
\begin{eqnarray}
\frac{\rmd}{\rmd t} \int_{S}\rmd^2x\,  \bnu\cdot\bi{B}
&=& \int_{S}\rmd^2x \, \bnu\cdot(\partial_t \bi{B})
\nonumber \\
&=& \oint_{\partial S}\rmd x \, \btau\cdot(\partial_t \bi{A}) 
\nonumber \\
&=& - \oint_{\partial S}\rmd x \, \btau\cdot(\bi{E}+\nabla\phi) 
=0,
\label{flux_conservation}
\end{eqnarray}
since $\btau\cdot\bi{E}=0$ by (\ref{BC-E}), and $\nabla\phi$ is an exact differential.
Assuming that $\bnu\cdot\bi{B}=0$ at $t=0$, we obtain the homogeneous boundary condition (\ref{BC-B}).

\subsection{Total flux conservation: cohomology constraint}  %--------------------------------
\label{subsec:cohomology}

When the domain $\Omega$ is multiply connected, the boundary conditions (\ref{BC-V}) and (\ref{BC-B}) are
insufficient to determine a unique solution; we have to specify the ``magnetic flux''
on each \emph{cut} $\Sigma_\ell$ of the handle of $\Omega$.
Here, we make 
$\Omega$ into a simply connected domain $\Omega_0$ by inserting cuts $\Sigma_\ell$ across each handle of $\Omega$:
$\Omega_0:=\Omega\setminus(\bigcup_{\ell=1}^m \Sigma_\ell)$, %to be a simply connected domain
where $m$ is  the \emph{genus} of $\Omega$ (see  \,\ref{appendix:cohomology}).

Hereafter, we assume $m\geq1$.
The \emph{fluxes} of $\bi{B}$, given by 
\begin{equation}
\Phi_\ell(\bi{B}) = \int_{\Sigma_\ell}\rmd^2x \, \bnu\cdot\bi{B}
\quad (\ell=1,\cdots,m),
\label{flux}
\end{equation}
are the constants of motion ($\bnu$ is the unit normal vector of $\Sigma_\ell$)
when we assume the perfectly conducting boundary condition (\ref{BC-E}). 
In fact, replacing $S$ by $\Sigma_\ell$ in (\ref{flux_conservation}), 
we obtain $\rmd\Phi_\ell/\rmd t=0$, since the boundary 
$\partial\Sigma_\ell$ of $\Sigma_\ell$ is a cycle on $\partial\Omega$ where the
tangential electric field vanishes.

The flux conditions $\Phi_\ell(\bi{B}) =$ constant ($\ell=1,\cdots,m$)
mean that the cohomology class of 2-forms ($\bi{B}_H$ such that
$\nabla\times\bi{B}_H=0$, $\nabla\cdot\bi{B}_H=0$, $\bnu\cdot\bi{B}_H|_{\partial\Omega}=0$)
included in $\bi{B}$ are fixed constants (see  \,\ref{appendix:cohomology}).

\subsection{Local flux conservation and circulation theorem}  %------------------------------
\label{subsec:circulation}

Whereas the aforementioned magnetic flux constraints pertain to the cohomology of the fixed domain $\Omega$
(which  restrict a finite number $m$ degrees of freedom),
every local magnetic flux on  an arbitrary co-moving surface $\sigma$ is also constrained,
i.e.,  the magnetic flux (or, equivalently, the circulation of the vector potential along the boundary
$\partial\sigma$ of the disk $\sigma$)
\[
\Phi_\sigma(t) = \int_{\sigma(t)} \rmd^2 x\, \bnu\cdot\bi{B} 
=\oint_{\partial\sigma(t)} \rmd x\, \btau\cdot\bi{A} 
\]
is a constant of motion.
This conservation law (often called Alfv\'en's theorem in the MHD context, but  equivalent to Kelvin's circulation theorem) 
is a direct consequence of the magnetic induction equation (\ref{MHD-induction}),
which implies that the 2-form $\bi{B}$ is Lie-dragged by the flow $\bi{V}$.
Because of  this infinite set of conservation laws, the magnetic field lines are forbidden to change their  topology.

In the next section, we will study the meaning of these total and local flux conservation laws from  the perspective of Hamiltonian mechanics.

%%%%%%%%%%%%%%%%%%%%%%%%%%%%%%%%%%%%%%%%%%%%%%%%%%%%%%%%%%%%%%%%%%%%%%%%%%%%%%%%%%%%%%%%%%%%%%%%%%
%%%%%%%%%%%%%%%%%%%%%%%%%%%%%%%%%%%%%%%%%%%%%%%%%%%%%%%%%%%%%%%%%%%%%%%%%%%%%%%%%%%%%%%%%%%%%%%%%%
%%%%%%%%%%%%%%%%%%%%%%%%%%%%%%%%%%%%%%%%%%%%%%%%%%%%%%%%%%%%%%%%%%%%%%%%%%%%%%%%%%%%%%%%%%%%%%%%%%
%%%%%%%%%%%%%%%%%%%%%%%%%%%%%%%%%%%%%%%%%%%%%%%%%%%%%%%%%%%%%%%%%%%%%%%%%%%%%%%%%%%%%%%%%%%%%%%%%%
%%%%%%%%%%%%%%%%%%%%%%%%%%%%%%%%%%%%%%%%%%%%%%%%%%%%%%%%%%%%%%%%%%%%%%%%%%%%%%%%%%%%%%%%%%%%%%%%%%
\section{Hamiltonian structure  of magnetohydrodynamics}
\label{sec:Hamiltonian}

\subsection{Noncanonical Poisson bracket and Casimir elements}  %-------------------------------

The foregoing MHD equations  possesses the noncanonical Hamiltonian form first given in \cite{MG80}, where   
the  phase space $X$ contains the state vector $u=~^t(\rho,\bi{V},\bi{B})$, 
and the  Hamiltonian and  Poisson operator are given as follows:
\begin{eqnarray}
H &=& \int_\Omega \rmd^3x \, \left\{ {\rho}\left[ \frac{V^2}{2} 
+  \mathcal{E}({\rho})\right]+ \frac{B^2}{2} \right\},
\label{MHD-Hamiltonian}
\\
\mathcal{J} &=& {\small \left( \begin{array}{ccc}
0 & -\nabla\cdot & 0
\\
-\nabla & -{\rho}^{-1}(\nabla\times\bi{V})\times 
& {\rho}^{-1}(\nabla\times\circ)\times\bi{B}
\\
0 & \nabla\times\left( \circ\times{\rho}^{-1}\bi{B} \right) & 0
\end{array} \right) } . 
\label{MHD-J}
\end{eqnarray}
Here $\circ$ implies insertion of the function to the right of the operator.
We formally endow the phase space $X$ with the standard $L^2$ norm.
The Poisson operator $\mathcal{J}$ is a differential operator with inhomogeneous coefficients, and the
 domain of $\mathcal{J}$ is a subspace of $X$ such that
\begin{equation}
% D(\mathcal{J}) = \{ ~^t(\rho^\dagger,\bi{V}^\dagger,\bi{B}^\dagger)\in H^2(\Omega);\,
D(\mathcal{J}) = \{ ~^t(\rho^\dagger,\bi{V}^\dagger,\bi{B}^\dagger);\,
\bnu\cdot\bi{V}^\dagger=\bnu\cdot\bi{B}^\dagger=0, \nabla\cdot\bi{B}^\dagger=0 \}.
\end{equation}
There are subtleties associated with the   mathematical identification of $D(\mathcal{J})$ and we  will  address the minimum amount needed for our purposes here.

% We note that a general state vector may not reside in the rather narrow subspace $D(\mathcal{J})$
% (in fact, the existence theorem of the solution to the nonlinear system like MHD is not known).
% When we study singularities, we evaluate the Poisson bracket $[F,G]$ in a weak sense.
It is easily  verified that a Poisson bracket $[ F, G] = \langle \partial_{u} F, \mathcal{J}\partial_u G\rangle$ is antisymmetric and using the techniques of \cite{aip82} it was verified that it satisfies Jacobi's identity.
When the specific enthalpy $h(\rho)=\partial (\rho\mathcal{E}(\rho))/\partial\rho$ is a
continuous function, $H(\rho,\bi{V},\bi{B})$ is a $C^1$-class functional of the state vector $u=~^t(\rho,\bi{V},\bi{B})$, and the functional gradient $\partial_u H(u)$ is evaluated in the classical sense.
With this structure, the Hamilton   form of (\ref{Hamilton_eq_2}) reproduces the MHD equations (\ref{MHD-mass_conservation})-(\ref{MHD-induction}).

The Poisson operator $\mathcal{J}$ has well-known Casimir elements \cite{aip82,padhye,Hameiri2004,amp1}:
\begin{eqnarray}
C_1 &=& \int_\Omega \rmd^3x\, {\rho} ,
\label{Casimir-mass}
\\
C_2 &=& \frac{1}{2} \int_\Omega \rmd^3x\, \bi{A}\cdot\bi{B} ,
\label{CasimirM} 
\\
C_3 &=& \int_\Omega \rmd^3x\, \bi{V}\cdot\bi{B}  ,
\label{CasimirK} 
\end{eqnarray}
where $\bi{A}$ is the vector potential ($\bi{A}=\mathrm{curl}^{-1}\bi{B}$),
which is evaluated with a fixed gauge and boundary conditions.
The Casimir $C_1$ is the total mass,
$C_2$ the magnetic helicity, and $C_3$ the cross helicity.

\subsection{A Casimir element representing the total fluxes} %----------------------------

The total flux pertinent to the cohomology of the domain $\Omega$ 
can be regarded as a \emph{singular} Casimir element of the MHD system.
We may formally write
\[
\Phi_\ell(\bi{B}) = \int_{\Sigma_\ell} \rmd^2x\, \bi{n}\cdot\bi{B}  
= \int_\Omega \rmd^3x\, \bsigma_\ell\cdot\bi{B}
\]
with a singular 1-form such that
\[
\bsigma_\ell=\nabla \lshad\bar{\theta}_\ell\rshad,
% \bi{a}=\nabla \lshad \theta_\ell \rshad.
\]
where $\bar{\theta}_\ell=\theta_\ell/(2\pi)$ with 
$\theta_\ell$  the angle measured from $\Sigma_\ell$ going around the handle $\ell$,
and $\lshad\alpha\rshad$ is Gauss's symbol for the maximum integer smaller than $\alpha\in\mathbb{R}$,
i.e.,  $\lshad\bar{\theta}_\ell\rshad=\lshad\theta_\ell/(2\pi)\rshad$ is the ``winding number'' of the angle $\theta_\ell$,
which  steps by unity  at $\Sigma_\ell$ (see  \ref{appendix:cohomology}).
Formally, we calculate  $\partial_{u} \Phi_\ell(\bi{B}) = (0,0,\nabla \lshad\bar{\theta}_\ell\rshad)$, and
$\nabla\times(\nabla \lshad\bar{\theta}_\ell\rshad)=0$, 
hence, $\Phi_\ell(\bi{B})$ is a Casimir element.

\begin{remark}[separation of cohomology]
If the domain $\Omega$ is multiply connected (i.e.,  the genus $m\geq1$) 
and the magnetic flux $\Phi_\ell(\bi{B})$ on each handle ($\ell=1,\cdots,m$) is
constrained by the boundary condition (\ref{BC-E}),
only the internal magnetic field $\bi{B}_\Sigma=\bi{B}-\bi{B}_H$ is the dynamical variable 
(see \,\ref{appendix:cohomology}).
We may replace the total $\bi{B}$ by $\bi{B}_\Sigma$ in defining the state vector $u$.
Then, the Casimir elements $\Phi_1(\bi{B}_\Sigma),\cdots,\Phi_m(\bi{B}_\Sigma)$ trivialize,   
and we  define the magnetic helicity as
\begin{equation}
C'_2= \frac{1}{2} \int_\Omega \rmd^3 x \bi{A}_\Sigma\cdot\bi{B}_\Sigma +  
\int_\Omega \rmd^3 x \bi{A}_H\cdot\bi{B}_\Sigma,
\label{CasimirM'}
\end{equation}
where $\nabla\times\bi{A}_\Sigma=\bi{A}_\Sigma$ and $\bi{A}_H=\bi{B}_H$ 
(see Remark 1 of \cite{YD2012}).
% We may separate the harmonic component $\bi{B}_H$ from the total $\bi{B}$ (which is fixed by the flux condition),
% and define the phase space only by the dynamical component $\bi{B}_\Sigma=\bi{B}-\bi{B}_H$.
% In this paper, however, we deal the total $\bi{B}$ and explain the
% constancy of $\bi{B}_H$ by unearthing a Casimir element constraining the flux.
\end{remark}

\subsection{Extension of the phase space} %-------------------------------------

To formulate the local magnetic flux as a Casimir element, we  
extend  the phase space as in Sec.~\ref{subsec:2D_example}  in order to include topological indexes information in  the set of dynamical variables.
Adding a $2$-form $\check{\bi{B}}$,
which we call a  \emph{mock field}, to the MHD variables,
gives the  extended phase space  state vector
\begin{equation}
\tilde{u} = ~^t(\rho,\bi{V},\bi{B},\check{\bi{B}}),
\label{extension}
\end{equation}
on which we define a degenerate Poisson manifold by
\begin{equation}
\tilde{\mathcal{J}} = {\small \left( \begin{array}{cccc}
0 & -\nabla\cdot & 0 & 0
\\
-\nabla & -{\rho}^{-1}(\nabla\times\bi{V})\times 
& {\rho}^{-1}(\nabla\times\circ)\times\bi{B} & {\rho}^{-1}(\nabla\times\circ)\times\check{\bi{B}}
\\
0 & \nabla\times\left( \circ\times{\rho}^{-1}\bi{B} \right) & 0 & 0 
\\
0 & \nabla\times\left( \circ\times{\rho}^{-1}\check{\bi{B}} \right) & 0 & 0 
\end{array} \right) } .
\label{MHD-J-extended}
\end{equation}
We assume that $\check{\bi{B}}$ obeys the same boundary condition as $\bi{B}$, 
\begin{equation}
\bi{n}\cdot\check{\bi{B}}|_{\partial\Omega} = 0.
\label{BC-B2}
\end{equation}
Using the same Hamiltonian (\ref{MHD-Hamiltonian}), we obtain an extended dynamics
governed by exactly the same equations (\ref{MHD-mass_conservation})-(\ref{MHD-induction})
together with an additional equation
\begin{equation}
\partial_t \check{\bi{B}}=\nabla\times(\bi{V}\times\check{\bi{B}}) .
\label{MHD-induction2}
\end{equation}
The projection of the orbit onto the original phase space reproduces the same dynamics;
the mock field $\check{\bi{B}}$ is just a passive 2-covector moved by the flow $\bi{V}$
of the original system.

The extended Poisson operator (\ref{MHD-J-extended}) has the  set of Casimir elements composed of 
$C_1$, $C_2$, and a new \emph{cross helicity}
\begin{equation}
C_4 = \int_\Omega \rmd^3x\, \bi{A}\cdot\check{\bi{B}} ,
\label{CasimirC2} 
\end{equation}
as well as a mock magnetic helicity
\begin{equation}
C_5 = \frac{1}{2}\int_\Omega \rmd^3x\, \check{\bi{A}}\cdot\check{\bi{B}} .
\label{CasimirM2} 
\end{equation}

Interestingly, the original (standard) cross helicity 
$C_3 = \int_\Omega \rmd^3x\, \bi{V}\cdot\bi{B} $ is no longer a Casimir element of the extended system,
although it is still a constant of motion.
The constancy of $C_3$ is now due to the ``symmetry'' of a Hamiltonian
with ignorable dependence on  the mock field $\check{\bi{B}}$;
for \emph{every} Hamiltonian $H(\rho,\bi{V},\bi{B})$,
which is not necessarily the MHD Hamiltonian (\ref{MHD-Hamiltonian}),
we find,
denoting $\rho^\dagger=\partial_\rho H$,
$\bi{V}^\dagger=\partial_{\bi{V}} H$,
$\bi{B}^\dagger=\partial_{\bi{B}} H$,
and noticing $\partial_{\check{\bi{B}}} H=0$
(while $H$ may be an arbitrary $C^1$-class functional of $u$,
we must assume $\partial_{u}H=\,^t(\rho^\dagger,\bi{V}^\dagger,\bi{B}^\dagger)\in D(\mathcal{J})$),
\begin{eqnarray*}
\frac{\rmd}{\rmd t} C_3 &=&
\int_\Omega \rmd^3 x\, \left\{(\partial_t\bi{V})\cdot\bi{B} + \bi{V}\cdot(\partial_t\bi{B}) \right\}
\\
&=& \int_\Omega \rmd^3 x\, \left\{ \left[
-\nabla\rho^\dagger - \rho^{-1}(\nabla\times\bi{V})\times\bi{V}^\dagger + \rho^{-1}(\nabla\times\bi{B}^\dagger)\times\bi{B}\right]\cdot\bi{B} \right.
\\
& & ~~~~\left. + \bi{V}\cdot\left[\nabla\times(\rho^{-1}\bi{V}^\dagger\times\bi{B})\right]\right\} 
\\
% &=& \int_{\partial\Omega}\bnu\cdot[(\rho^{-1}\bi{V}^\dagger\times\bi{B})\times\bi{V}]\,\rmd^2 x.
&=& 0.
\end{eqnarray*}
Here we have used the boundary condition $\bnu\cdot\bi{V}^\dagger=0$, which is
guaranteed for $\partial_{u}H\in D(\mathcal{J})$.

% \textcolor{red}{
% \begin{remark}[singular Casimirs]
% Obviously, the ``rank'' of $\tilde{\mathcal{J}}$ reduced to that of $\mathcal{J}$
% at $\check{\bi{B}}=\alpha\bi{B}$ ($\forall\alpha\in\mathbb{R}$).
% Singular Casimirs stem from this singularity.
% Whereas the original cross helicity 
% $C_3=\int_\Omega \rmd^3x\,\bi{V}\cdot\bi{B}$ is no longer a Casimir
% of $\tilde{\mathcal{J}}$, it is recovered as an \emph{interior Casimir}
% on a submanifold $M_\alpha=\{\tilde{u}=~^t(\rho,\bi{V},\bi{B},\alpha{\bi{B}})\}$
% ($\alpha\in\mathbb{R}$).
% Evidently, $M_\alpha$ is a coadjoint orbit (i.e. if $\check{\bi{B}}=\alpha\bi{B}$ at $t=0$,
% $\check{\bi{B}}=\alpha\bi{B}$ for all $t$),
% which is a leaf of the following singular Casimir:
% \begin{equation}
% C_{\alpha} = Y(\|{\bi{B}}_\alpha \|^2),
% \label{exterior_Casimir}
% \end{equation}
% where $Y(\tau)$ is the Heaviside function and ${\bi{B}}_\alpha=\alpha\bi{B}-\check{\bi{B}}$.
% Formally, we may calculate as
% \[
% \tilde{\mathcal{J}}\partial_{\tilde{u}}C_{\alpha} 
% = ~^t(0,2\alpha\delta(\|\bi{B}_\alpha\|^2)\rho^{-1} (\nabla \times\bi{B}_\alpha) \times{\bi{B}}_\alpha,0,0) =0.
% \]
% While these \emph{exterior} Casimirs are ``visible'' in the extended phase space,
% the interior Casimir is not; $C_3$ is a Casimir element only on the singularity
% $\bi{B}_\alpha=0$, which is the leaf of $C_{ex,\alpha}$
% (notice that the leaf of $C_{\alpha}$ exists only at the singularity).
% Projecting $\tilde{\mathcal{J}}$ onto the singular leaf $M_\alpha$
% recovers the original $\mathcal{J}$, and, of course, $C_3$ is a Casimir of $\mathcal{J}$.
% \end{remark}
% }

\subsection{Local flux (circulation) as a Casimir element} %-------------------------------

Here we show that the cross helicity $C_4$ is the  \emph{circulation} of $\bi{A}$
for a ``pure-state 2-form'' $\check{\bi{B}}$. 
We can consider a \emph{filamentary} $\check{\bi{B}}$ supported on a co-moving loop $L(t)$ 
such that, for every disk $\sigma$,
\begin{equation}
\int_\sigma \rmd^2 x\, \bnu\cdot\check{\bi{B}}  = 
\mathcal{L}(L(t),\partial\sigma),
% \left\{ \begin{array}{ll}
% 1 & \textrm{ if } L \cap \sigma \neq \emptyset ,
% \\
% 0 & \textrm{ if } L \cap \sigma = \emptyset .
% \end{array} \right.
\label{B-filament-0}
\end{equation}
where $\mathcal{L}(L_1,L_2)$ denotes the linking number of two loops $L_1$ and $L_2$
(the exact definition will be given in Sec.\,\ref{sec:filament}).
Formally, the filamentary $\check{\bi{B}}$
is a delta-measure on a co-moving loop $L(t)$ carrying a unit mock flux.
Inserting such $\check{\bi{B}}$ into the cross helicity $C_4$, we obtain
\begin{equation}
C_4 = \int_\Omega \rmd^3x\, \bi{A}\cdot\check{\bi{B}} 
= \oint_{L(t)} \rmd x \, \btau\cdot\bi{A}.
\label{CasimirC2-circulation} 
\end{equation}
Hence, the conservation of the cross helicity $C_4$ implies the conservation of the circulation,
or equivalently, the local magnetic flux conservation on every disk bounded by $L(t)$.

\begin{remark}[two-dimensional MHD system]
\label{remark:2D}
In the  two-dimensional system of Sec.\,\ref{subsec:2D_example}, the cross helicity $C_4$ parallels the Casimir element $\int\rmd^2 x f(\psi)$ of
the reduced MHD system (see\cite{aip82,MH84,MM84,Morrison1998}).
To see this  consider a cylindrical domain $\Omega=\Sigma\times[0,1]$
($\Sigma\subset\mathbb{R}^2$) and two-dimensional vectors
$\bi{V}=\,^t(\partial_y\varphi,-\partial_x\varphi)$ and
$\bi{B}= \,^t(\partial_y\psi,-\partial_x\psi)$, which
satisfy periodic boundary conditions at $z=0$ and $L$.
We may assume $\bi{A}=\psi\bi{e}_z$.
With a constant $\rho$, $u=\,^t(\rho,\bi{V},\bi{B})$ may satisfy the MHD equations
(\ref{MHD-mass_conservation})-(\ref{MHD-induction}) in $\Omega$,
as well as the reduced MHD equations, the system (II) of Sec.\,\ref{subsec:2D_example}, in $\Sigma$.
% Let $\sigma(t)\subset\Sigma$ be a co-moving surface, and $\check{\bi{B}}=\varrho(\sigma(t))\bi{e}_z$,
% where $\varrho(\sigma)$ is the characteristic function of $\sigma$.
Let $\bxi(t)$ be a co-moving point in $\Sigma$, 
and $\check{\bi{B}}=\delta(\bi{x}-\bxi(t))\bi{e}_z$.
Then, 
\[
C_4 = \int \rmd^2 z \delta(\bi{x}-\bxi(t))\psi(\bi{x}) = \psi(\bxi(t)).
\]
Integrating $C_4$ over all points $\bxi(0)\in\Sigma$ with a weight function $f$ yields
$\int\rmd^2 x f(\psi)$.
\end{remark}

In the next section, we shall identify the unit-flux filament as a pure states of
a Banach algebra, and show that the co-moving filament is a 
singular solution of (\ref{MHD-induction2}).

%%%%%%%%%%%%%%%%%%%%%%%%%%%%%%%%%%%%%%%%%%%%%%%%%%%%%%%%%%%%%%%%%%%%%%%%%%%%%%%%%%%%%%%%%%%%%%%%%%
%%%%%%%%%%%%%%%%%%%%%%%%%%%%%%%%%%%%%%%%%%%%%%%%%%%%%%%%%%%%%%%%%%%%%%%%%%%%%%%%%%%%%%%%%%%%%%%%%%
%%%%%%%%%%%%%%%%%%%%%%%%%%%%%%%%%%%%%%%%%%%%%%%%%%%%%%%%%%%%%%%%%%%%%%%%%%%%%%%%%%%%%%%%%%%%%%%%%%
%%%%%%%%%%%%%%%%%%%%%%%%%%%%%%%%%%%%%%%%%%%%%%%%%%%%%%%%%%%%%%%%%%%%%%%%%%%%%%%%%%%%%%%%%%%%%%%%%%
%%%%%%%%%%%%%%%%%%%%%%%%%%%%%%%%%%%%%%%%%%%%%%%%%%%%%%%%%%%%%%%%%%%%%%%%%%%%%%%%%%%%%%%%%%%%%%%%%%
\section{Dynamics of loops: Poincar\'e dual of local flux}
\label{sec:filament}

\subsection{Pure state of Banach algebra} %-------------------------------------------------------
\label{subsec:pure-state}

A unit-flux filament is identified as
a \emph{pure-state} 2-form (physically a vorticity or a magnetic field, 
which, however, is a mock field)\,\cite{YKY2013}.
Naturally, a 2-form is in the Poincar\'e-dual relation with a 2-chain
(two-dimensional surface), and a pure-state 2-form is a 2-dimensional surface measure.
The filamentary $\check{\bi{B}}$ is, then, the temporal cross-section of a 2-chain in the space-time.

\begin{definition}[pure sate]
\label{definition:pure_state}
Let $M$ be a smooth manifold of dimension $n$,
and $\Omega\subset M$ be a $p$-dimensional connected null-boundary submanifold of class $C^1$.
Each $\Omega$ can be regarded as an equivalent of a 
\emph{pure-sate functional} $\eta_\Omega$ on the space $\wedge^p T^*M$ of continuous $p$-forms:
\[
\eta_\Omega:\, \omega %= \langle \omega|J(\Omega)\rangle
% = \int \omega\wedge J(\Omega)
\mapsto \int_\Omega \omega,
% \quad (\forall\omega\in\wedge^p T^*M),
\]
which can be represented as
\[
\eta_\Omega(\omega) %= \langle \omega|J(\Omega)\rangle
= \int_M \mathfrak{J}(\Omega)\wedge\omega
= \int_\Omega \omega
\]
with an $(n-p)$-dimensional $\delta$-measure 
$\mathfrak{J}(\Omega)=\wedge^{n-p}\delta(x^\mu-\xi^\mu)\rmd x^\mu $,
where $x^\mu$ are local coordinates, and 
\[
\mathrm{supp}\,\mathfrak{J}(\Omega) = \Omega = \{ \bi{x}\in\mathbb{R}^n;\, x^\mu=\xi^\mu \,(\mu=1,\cdots,n-p) \} .
\]
We call $\mathfrak{J}(\Omega)$ a \emph{pure state} $(n-p)$-form,
which is a member of the Hodge-dual space of $\wedge^p T^*M$.
\end{definition}

% On compact manifolds, the pure state of a submanifold $\Omega$ 
% is obtained as a limit of the Thom form of $\Omega$,
% and the cohomology class of a pure state of $\Omega$ 
% corresponds to the Poincar\'e dual of $\Omega$\,\cite{Poincare-dual}.
% We also note that the duality between submanifolds and pure state parallels the duality 
% between infinite chains in homologies with closed support (or Borel-Moore homology)
% and cochains in cohomologies\,\cite{cochain}.

\subsection{Orbit of a filament} %-------------------------------------------------------
\label{subsec:orbit}

Here we show that the co-moving pure-state filament is a (singular)
solution of (\ref{MHD-induction2}).
For the convenience of formulation, we rewrite the determining equation
(\ref{MHD-induction2}) of the mock field $\check{\bi{B}}$ in the four-dimensional
Galilei space-time $M_G$
(we draw heavily on the theory of relativistic helicity in Minkowski space-time developed in
\cite{YKY2013}).
Normalizing the speed of light so $c=1$, we denote the four-dimensional coordinates as
$(x^0,x^1,x^2.x^3)=(t,x,y,z) \in M_G$.
The (nonrelativistic) four-vector is
$U=U^\mu\partial_\mu = (\rmd x^\mu/\rmd t)\partial_\mu \in TM_G$,
which has four components $U= (1,\bi{V})$.
We may identify the mock field $\check{\bi{B}}$ as the three-vector part of a 2-form:
we define $F=F_{\mu\nu}\rmd x^\mu\wedge\rmd x^\nu/2$ with
a ``Faraday tensor''
% denoting the EM four potential (1-form) by $A = A_\mu \rmd x^\mu$,
% the Faraday tensor is 
% $F=\rmd A = \partial_\mu A_\nu \rmd x^\mu\wedge\rmd x^\nu = F_{\mu\nu}\rmd x^\mu\wedge\rmd x^\nu/2$,
% where
% =-\nabla A_0 -\partial_t\bi{A}$ and $\check{\bi{B}} = (\nabla\times\bi{A})$ ($\bi{A}=(A_1,A_2,A_3)$).
\begin{equation}
F_{\mu\nu} 
= \left( \begin{array}{cccc}
 0             &  \check{E}_1 &  \check{E}_2 &  \check{E}_3 \\
-\check{E}_1 &  0             & -\check{B}_3 &  \check{B}_2 \\
-\check{E}_2 &  \check{B}_3 &  0             & -\check{B}_1 \\
-\check{E}_3 & -\check{B}_2 &  \check{B}_1 &  0             \\
\end{array} \right) ,
\label{M-components_down}
\end{equation}
where $\check{\bi{E}}$ is a certain three-vector satisfying Faraday's law
\begin{equation}
\nabla\times\check{\bi{E}} = - \partial_t \check{\bi{B}}.
\label{Faraday}
\end{equation}
Invoking these notations, the ``vorticity equation'' (\ref{MHD-induction2}) reads
\begin{equation}
\rmd i_U F = 0.
\label{vorticity_equation_diff-form}
\end{equation}
By (\ref{Faraday}) together with $\nabla\cdot\check{\bi{B}}=0$, $F$ is a closed 2-form ($\rmd F=0$),
thus we may rewrite (\ref{vorticity_equation_diff-form}) as
\begin{equation}
L_U F = 0,
\label{vorticity_equation_diff-form-2}
\end{equation}
where $L_U = \rmd i_U + i_U \rmd$ is the Lie derivative.
Notice that (\ref{vorticity_equation_diff-form}) consists of six independent equations;
three of them are (\ref{MHD-induction2}), and
the others are the energy equation
\[
\partial_t (\bi{E}+\bi{V}\times\bi{B})-\nabla(\bi{E}\cdot\bi{V}) = 0,
\]
which is solved by a potential energy $\phi$ such that $\bi{E}\cdot\bi{V}=-\partial_t\phi$
and $\bi{E}+\bi{V}\times\bi{B}=-\nabla\phi$.

Let $\mathfrak{J}(\Gamma_0)$ be a pure-state 3-form (vortex filament) 
supported on a loop $\Gamma_0$ in $M_G$,
which we may write
\begin{eqnarray}
\mathfrak{J}(\Gamma_0) = \delta_{\Gamma_0} b 
&=& \delta_{\Gamma_0} (b_1 \rmd x^0\wedge\rmd x^2\wedge \rmd x^3
\nonumber \\
& & ~~~~- b_2 \rmd x^0\wedge\rmd x^1\wedge \rmd x^3
+ b_3 \rmd x^0\wedge\rmd x^1\wedge \rmd x^2).
\label{pure-B}
\end{eqnarray}
We denote by $\mathcal{T}_U(t)$ the diffeomorphism generated by the vector $U$
(i.e. $\rmd \mathcal{T}_U(t)/\rmd t = U$).
The orbit of $\Gamma(t) = \mathcal{T}_U(t)\Gamma_0$ defines a surface (2-chain)
\begin{equation}
\Sigma = \bigcup_{t\in\mathbb{R}} \Gamma(t),
\label{vortex_sheet}
\end{equation}
and its Poincar\'e-dual is written as
\begin{equation}
\mathfrak{J}(\Sigma) = - \delta_\Sigma i_U b = \delta_\Sigma  \frac{1}{2} \mathcal{F}_{\mu\nu} \rmd x^\mu\wedge\rmd x^\nu,
\label{vortex_sheet-dual}
\end{equation}
where 
% $b= b_1 \rmd x^0\wedge\rmd x^2\wedge \rmd x^3
% - b_2 \rmd x^0\wedge\rmd x^1\wedge \rmd x^3
% + b_3 \rmd x^0\wedge\rmd x^1\wedge \rmd x^2$ and
\begin{equation}
\mathcal{F}_{\mu\nu} 
= \left( \begin{array}{cccc}
 0             &  \varepsilon_1 &  \varepsilon_2 &  \varepsilon_3 \\
-\varepsilon_1 &  0             & -{b}_3 &  {b}_2 \\
-\varepsilon_2 &  {b}_3 &  0             & -{b}_1 \\
-\varepsilon_3 & -{b}_2 &  {b}_1 &  0             \\
\end{array} \right) 
\label{F-components}
\end{equation}
with $\bvarepsilon = - \bi{V}\times\bi{b} $.
Evidently,
\[
i_U \mathfrak{J}(\Sigma) =  - \delta_\Sigma i_U i_U b =0; 
\]
hence,  $F=\mathfrak{J}(\Sigma)$ satisfies (\ref{vorticity_equation_diff-form}).
The $t$-plane projection of $\mathfrak{J}(\Sigma)$ yields
(denoting $\Xi(t)=\{x;\, x^0=t\}$) 
\begin{equation}
- \delta_{\Xi(t)} \rmd x^0\wedge \mathfrak{J}(\Sigma)
= \delta_{\Gamma(t)} b(t),
\label{projection}
\end{equation}
which is a pure-state filament on a co-moving loop $\Gamma(t)$.
Now we have

\begin{theorem}
\label{theorem:dynamics}
Suppose that an initial mock field $\check{\bi{B}}(0)$ is given as a pure-state
$\mathfrak{J}(\Gamma(0))$ on a loop $\Gamma(0)$ bounding a disk.  Then, the orbit 
$\Sigma = \cup_{t\in\mathbb{R}} \Gamma(t)$ defines a pure-state 2-from
$\mathfrak{J}(\Sigma)$ that satisfies the vorticity equation (\ref{vorticity_equation_diff-form}).
The $t$-plane projection of $\mathfrak{J}(\Sigma)$ is 
a pure-state filament $\check{\bi{B}}(t)=\mathfrak{J}(\Gamma(t))$ on a co-moving loop $\Gamma(t)$.
\end{theorem}

%%%%%%%%%%%%%%%%%%%%%%%%%%%%%%%%%%%%%%%%%%%%%%%%%%%%%%%%%%%%%%%%%%%%%%%%%%%%%%%%%%%%%%%%%%%%%%%%%%
%%%%%%%%%%%%%%%%%%%%%%%%%%%%%%%%%%%%%%%%%%%%%%%%%%%%%%%%%%%%%%%%%%%%%%%%%%%%%%%%%%%%%%%%%%%%%%%%%%
%%%%%%%%%%%%%%%%%%%%%%%%%%%%%%%%%%%%%%%%%%%%%%%%%%%%%%%%%%%%%%%%%%%%%%%%%%%%%%%%%%%%%%%%%%%%%%%%%%
%%%%%%%%%%%%%%%%%%%%%%%%%%%%%%%%%%%%%%%%%%%%%%%%%%%%%%%%%%%%%%%%%%%%%%%%%%%%%%%%%%%%%%%%%%%%%%%%%%
%%%%%%%%%%%%%%%%%%%%%%%%%%%%%%%%%%%%%%%%%%%%%%%%%%%%%%%%%%%%%%%%%%%%%%%%%%%%%%%%%%%%%%%%%%%%%%%%%%
\section{Singular Casimir element: application to tearing modes}
\label{sec:singular_Casimir}
In this section, we study a different type of singular Casimir element, 
the cross helicity of extended MHD,
which controls bifurcation of topologically different equilibria.
The theory is applied to the tearing modes that are bifurcated equilibria 
on Casimir leaves\,\cite{YD2012};
as long as the Casimir element is constrained, each tearing mode is stationary.
However, by a singular perturbation that unfreezes the Casimir element,
some tearing modes that have lower energies can be excited  by changing the cross helicity.

\subsection{Equilibrium points of energy-Casimir functional}  %---------------------------------
\label{subsec:energy-Casimir}
We start by reviewing the equilibria of standard energy-Casimir functionals.
When we have a Casimir element $C(u)$ in a noncanonical Hamiltonian system,
a transformation of the Hamiltonian $H(u)$ such as
\begin{equation}
H(u) \mapsto {H}_{{\mu}}(u)
= {H}(u) - \mu {C} (u)
% \quad (\mu\in\mathbb{R})
\label{Hamiltonian-system-3}
\end{equation}
(with an arbitrary real constant $\mu$) does not change the dynamics.
In fact, the Hamilton form  is invariant under this transformation.
We call the transformed Hamiltonian 
${H}_{{\mu}}(u)$ an  \emph{energy-Casimir} function\,
\cite{Kruskal1958,Hazeltine1984,Morrison86,Morrison1998,Arnold-Khesin}.

Interpreting the parameter $\mu$ as a Lagrange multiplier of the equilibrium variational principle,
${H}_{{\mu}}(u)$ is the effective Hamiltonian with the constraint that restricts the
Casimir element $C(u)$
to be a given value (since $C(u)$ is a constant of motion, its value is fixed by its
initial value).
As we will see in some examples, Hamiltonians are rather simple, often being ``norms'' on  the phase space.
However, an energy-Casimir functional may have a nontrivial structure.
Geometrically, ${H}_{{\mu}}(u)$ is the distribution of $H(u)$ on a 
Casimir leaf.  If Casimir leaves are
distorted with respect to the energy norm, the effective Hamiltonian 
may have a complex distribution on the leaf, which is, in fact, the origin of
various interesting structures in noncanonical Hamiltonian systems.

Applying this energy-Casimir method to the MHD system, we obtain the Beltrami-Bernoulli equilibria,
constraining the Casimir elements (\ref{Casimir-mass})-(\ref{CasimirK}) on the Hamiltonian (\ref{MHD-Hamiltonian}), we consider
\begin{equation}
\partial_{u} H_{\mu_1,\mu_2,\mu_3} = 0,
\quad H_{\mu_1,\mu_2,\mu_3} = H-\mu_1 C_1 -\mu_2 C_2 - \mu_3 C_3 ,
\label{Beltrami-0}
\end{equation}
which reads as
\begin{eqnarray}
V^2/2 + h -\mu_1 =0,
\label{Beltrami-1} \\
\rho\bi{V}-\mu_3\bi{B} = 0,
% \bi{V}=0,
\label{Beltrami-2} \\
\nabla\times\bi{B} - \mu_2\bi{B} -\mu_3\nabla\times\bi{V} =0.
\label{Beltrami-3}
\end{eqnarray}
% where $h=\partial[\rho\mathcal{E}(\rho)]/\partial \rho$ is the specific enthalpy.
In deriving (\ref{Beltrami-3}), we have applied the curl operator. 
Putting $\mu_3=0$ simplifies the solutions to be the Beltrami fields such that
$\nabla\times\bi{B}=\mu_2\bi{B}$, $\bi{V}=0$, and $h=\mu_1$.

In the next subsection, we apply the energy-Casimir method to the extended MHD
system with a mock field $\check{\bi{B}}$,
and show that an interesting bifurcation occurs at a ``singularity'' in the phase space.

\subsection{Singular Casimir element}  %--------------------------------------------------------
\label{subsec:singular_Casimir}

Let us recall the determining equation of the cross helicity;
a functional $C(\bi{B},\check{\bi{B}})$ is a Casimir element if
\begin{equation}
\tilde{\mathcal{J}}\partial_{\tilde{u}} C(\bi{B},\check{\bi{B}})
= \,^t(0,\rho^{-1}[(\nabla\times\partial_{\bi{B}} C)\times\bi{B} +(\nabla\times\partial_{\check{\bi{B}}} C)\times\check{\bi{B}}], 0,0)
\label{cross-helicity-1}
\end{equation}
vanishes.
% where $\check{\bi{a}}=\partial_{\bi{B}} C$ and $\bi{a}=\partial_{\check{\bi{B}}} C$.
Evidently, $C_4=\int\rmd^3x \bi{A}\cdot\check{\bi{B}}$ (with arbitrary
$\bi{B}=\nabla\times\bi{A}$ and $\check{\bi{B}}=\nabla\times\check{\bi{A}}$) 
satisfies (\ref{cross-helicity-1}).
Here, we are interested in the \emph{singularity} at which the rank of $\tilde{\mathcal{J}}$ drops;
there is a pair of $\bi{B}^*=\nabla\times\bi{A}^*$ and 
$\check{\bi{B}}^*=\nabla\times\check{\bi{A}}^*$
such that the two terms on the right-hand side of (\ref{cross-helicity-1})
vanish separately, i.e.
\begin{equation}
% \bi{B}\times\check{\bi{B}} =
\bi{B}^*\times(\nabla\times\check{\bi{A}}^*) 
= (\nabla\times\bi{A}^*)\times\check{\bi{B}}^*
= 0.
\label{cross-helicity-2}
\end{equation}
% For given $\bi{B}$, we may solve (\ref{cross-helicity-2}) to determine $\check{\bi{A}}^*$.
We let $C_4^*:=C_4(\bi{B}^*,\check{\bi{B}}^*)$ and call it a \emph{singular cross helicity}.
A significance of $C_4^*$ is  (in addition to $\tilde{\mathcal{J}}\partial_{\tilde{u}} C_4^* =0$)
\[
\mathcal{J}\partial_{u} C_4^* =0 .
\]
% where the mock field $\check{\bi{B}}^*$ behaves as a parameter.

A trivial solution of (\ref{cross-helicity-2}) is $\bi{A}^*=\check{\bi{A}}^*$,
by which $C_4^*$ coincides with $C_2$,
i.e.,  the intersection of a $C_4$-leaf and a $C_2$-leaf is the singularity
(indeed, at the intersection of leaves, the rank of $\tilde{\mathcal{J}}$ must change). 

Interestingly, we may find nontrivial, hyperfunction solutions emerging from
the \emph{resonance singularity} of the differential equation (\ref{cross-helicity-2}).
Here we solve (\ref{cross-helicity-2}) for $\check{\bi{A}}^*$ by giving $\bi{B}^*$.
The determining equation can be rewritten as
\begin{equation}
\nabla\times\check{\bi{A}}^*= \eta\bi{B}^*
\label{linear-J-b'}
\end{equation}
with some scalar function $\eta$, which, however,
is not a free function;
the divergence of both sides of (\ref{linear-J-b'}) yields
\begin{equation}
\bi{B}^*\cdot\nabla\eta =0,
\label{linear-J-b-integrability}
\end{equation}
which implies that $\eta$ is constant along the magnetic field lines.
For the integrability of $\eta$, the magnetic field $\bi{B}^*$ must have 
integrable field lines; a continuous spatial symmetry guarantees this.
Here we consider a \emph{slab geometry}, in which we may write
$\bi{B}^* = \,^t\left( 0 , B^*_y(x) , B^*_z(x) \right)$.
% Each surface of $x=$ constant is a magnetic surface (integral surface of field lines).
% Denoting $\check{\bi{A}}^*=  \,^t\left( 0 , \check{A}^*_y(x), \check{A}^*_z(x) \right)$,
% (\ref{cross-helicity-2}) reads as
% \begin{equation}
% B_y\partial_x \check{A}^*_y + B_z\partial_x \check{A}^*_z =0,
% \label{slab_Casimir2}
% \end{equation}
% which may be solved for $\check{A}^*_y(x)$, given an arbitrary $\check{A}^*_z(x)$.
% Furthermore, we have \emph{singular} (hyper-function) solutions; 
Let us consider
\begin{equation}
\check{\bi{A}}^*=  \,^t\left( 0,  \check{A}^*_y(x), \check{A}^*_z(x)\right)  \rme^{\rmi(k_yy+k_zz)}.
\label{linear-J-b0}
\end{equation}
Putting $\check{A}^*_y(x)=\rmi k_y\vartheta(x)$ and $\check{A}^*_z(x)=\rmi k_z\vartheta(x)$, (\ref{cross-helicity-2}) reduces to
\begin{equation}
[B^*_y(x)k_y + B^*_z(x)k_z ]\partial_x \vartheta(x) = 0,
\label{linear-J-b1}
\end{equation}
% and $b_z(x)=(k_z/k_y)b_y(x)$, (\ref{linear-J-b}) holds
% (here we assume $k_y\neq0$; otherwise, we put $b_y=0$, and replace $b_y$ by $b_z$ in
% (\ref{linear-J-b1})).
which yields
\begin{equation}
\vartheta(x) = c_0 + c_1 Y(x-x_{r}),
\label{linear-J-b2}
\end{equation}
where % $Y(\cdot)$ is the  Heaviside step function, 
$c_0, \, c_1$ are complex constants, and
$k_y$, $k_z$ and $x_{r}$ (real constants) are chosen to
satisfy the \emph{resonance condition}
\begin{equation}
 B^*_y(x_{r})k_y + B^*_z(x_{r})k_z =0 .
\label{B-resonance}
\end{equation}
Then, 
$\eta = \rmi (k_y/B^*_z) \rme^{\rmi(k_yy+k_zz)}\delta(x-x_{r})$.

\begin{remark}[linear theory]
\label{remark:linear_tehory}
In the forgoing derivation, the singular Casimir element $C_4^*$ is essentially the same as 
the formulation of the \emph{resonant helical flux Casimir element} $C_b$ given in \cite{YD2012},
which was used to construct tearing modes.
It is remarkable, however, that the present argument is totally nonlinear,
while $C_b$ was formulated for a linearized Poisson operator (i.e., $\mathcal{J}(u)$ 
evaluated at an equilibrium point $u=u_0$).
These quantities  are compared as follows:
\begin{itemize}
\item
The singular cross helicity $C_4^*$ is a bilinear form combining the physical field $\bi{A}^*$ and the
mock field $\check{\bi{B}}^*$.
\item
The resonant helical flux is a linear form of 
a ``perturbation field'' $\tilde{\bi{B}}$ multiplied by a kernel element $\check{\bi{A}}^*$
(which is denoted by $\bi{b}$ in \cite{YD2012}) of $\mathcal{J}(u_0)$, 
\[
C_b(\tilde{\bi{B}}) = \int\rmd^3x \tilde{\bi{B}}\cdot\check{\bi{A}}^*.
\]
In the determining equation (\ref{cross-helicity-2}) of $\check{\bi{A}}^*$, 
$\bi{B}^*$ is regarded as an ``equilibrium field''.
\end{itemize}
By separating a perturbation $\tilde{\bi{B}}$ and an equilibrium $\bi{B}^*$,
$C_b$ is defined as a linear form on the space of perturbations. 
However, $C_4^*$ is a special value of $C_4$ evaluated at the singularity
$\bi{B}=\bi{B}^*=\nabla\times\bi{A}^*$ and $\check{\bi{B}}=\check{\bi{B}}^*$ in the phase space of total fields.
\end{remark}

\subsection{Tearing mode}  %--------------------------------------------------------
\label{subsec:tearing mode}

Because of the similarity between $C_4^*$ and the resonant helical flux Casimir element (see Remark\,\ref{remark:linear_tehory}),
the formulation of tearing modes goes almost parallel to that of \cite{YD2012} 
(see also \cite{YM2014}).
Here, we describe only the essence of the theory.

We begin with  an energy-Casimir functional on the extended phase space 
\begin{equation}
H_{\bmu} = H-\mu_1 C_1 -\mu_2 C_2 - \mu_4 C_4 -\mu_5 C_5 .
\label{energ-Casimir-extended}
\end{equation}
and consider a stationary point of 
\begin{equation}
\partial_{u} H_{\bmu} = 0.
\label{Beltrami-0-s}
\end{equation}
Notice that we are not demanding $\partial_{\tilde{u}} H_{\bmu} = 0$;
hence, the solution of (\ref{Beltrami-0-s}) is not necessarily an equilibrium point.
However, if we evaluate (\ref{Beltrami-0-s}) at the \emph{singularity} 
$\bi{B}=\bi{B}^*$ and $\check{\bi{B}}=\check{\bi{B}}^*$,
we obtain
\begin{equation}
\partial_{u} H_{\bmu}|_{\bi{B}^*,\check{\bi{B}}^*} 
=
\left( \begin{array}{c}
mV^2/2 + h -\mu_1
\\
\rho \bi{V}
\\
\bi{B}^* -\mu_2\bi{A}^* -\mu_4 \check{\bi{A}}^*
% \\
% -\mu_4 \bi{A}^* -\mu_5 \check{\bi{A}}^*
\end{array} \right)
= 0.
\label{Beltrami-1'}
\end{equation}
Let $M^*$ denote the solution of (\ref{Beltrami-1'}), then by 
 (\ref{cross-helicity-2}), we obtain
\[
\tilde{\mathcal{J}} \partial_{\tilde{u}} H_{\bmu} |_{M^*} 
= \tilde{\mathcal{J}} 
\left( \begin{array}{c}
0
\\
0
\\
0
\\
-\mu_4 \bi{A}^* -\mu_5 \check{\bi{A}}^*
\end{array} \right)
= 0.
\]
Hence, $M^*$ is an equilibrium of the extended MHD,
which \emph{bifurcates} from the singularity
$\bi{B}=\bi{B}^*$ and $\check{\bi{B}}=\check{\bi{B}}^*$.
By the determining equation (\ref{Beltrami-1'}), the equilibrium has zero velocity 
($\bi{V}=0$), constant enthalpy ($h=\mu_1$), and a magnetic field satisfying
\begin{equation}
% V^2/2 + h -\mu_1 =0,
% \label{EX-Beltrami-1} \\
% \bi{V}=0,
% \label{EX-Beltrami-2} \\
\nabla\times\bi{B}^* - \mu_2\bi{B}^*-\mu_4 \check{\bi{B}}^* =0,
\label{EX-Beltrami-3}
\end{equation}
where $\check{\bi{B}}^*$ is the hyper-function stemming from the resonant singularity.
The solution gives the tearing-mode equilibrium.

Note, as discussed above, we  can unfreeze the singular cross helicity $C_4^*$ 
by making a canonical pair with an angle variable (see Sec.\,\ref{subsec:canonization}).

%%%%%%%%%%%%%%%%%%%%%%%%%%%%%%%%%%%%%%%%%%%%%%%%%%%%%%%%%%%%%%%%%%%%%%%%%%%%%%%%%%%%%%%%%%%%%%%%%%
%%%%%%%%%%%%%%%%%%%%%%%%%%%%%%%%%%%%%%%%%%%%%%%%%%%%%%%%%%%%%%%%%%%%%%%%%%%%%%%%%%%%%%%%%%%%%%%%%%
%%%%%%%%%%%%%%%%%%%%%%%%%%%%%%%%%%%%%%%%%%%%%%%%%%%%%%%%%%%%%%%%%%%%%%%%%%%%%%%%%%%%%%%%%%%%%%%%%%
%%%%%%%%%%%%%%%%%%%%%%%%%%%%%%%%%%%%%%%%%%%%%%%%%%%%%%%%%%%%%%%%%%%%%%%%%%%%%%%%%%%%%%%%%%%%%%%%%%
%%%%%%%%%%%%%%%%%%%%%%%%%%%%%%%%%%%%%%%%%%%%%%%%%%%%%%%%%%%%%%%%%%%%%%%%%%%%%%%%%%%%%%%%%%%%%%%%%%
\section{Summary and conclusions}
\label{sec:conclusion}

By embedding a Poisson manifold of a noncanonical Hamiltonian system into  a higher-dimensional phase space, we can delineate topological structures within  a simpler picture.
For example,  we showed that the topological constraint on magnetic field lines in an ideal plasma,  Alfv'en's magnetic frozen-in law (or  vortex lines in a neutral fluid, Kelvin's circulation law), is not described by a foliation of the Poisson manifold because  these constraints are not \emph{integrable} to define Casimir leaves.  However, by 
introducing a \emph{mock field} and embedding the MHD system in a higher-dimensional phase space,  we found  that the local magnetic fluxes are represented as Casimir elements (cross helicities coupling the magnetic field and the mock field).
We have also elucidated the underlying Banach algebra describing the Poincar\'e duality of the
mock field and chains determining the local flux (or circulation).

The representation of a topological constraint by a Casimir element has an immense advantage
in studying the global structure of the Poisson manifold.
This is especially true for studying  \emph{singularities} on the manifold (where the rank of the Poisson operator changes)
where  different leaves intersect or new leaf bifurcations take place.
For a Poisson operator that is a differential operator, the singularity in the phase space (function space) is related to the singularity in the base space of the operator,
which yields singular (hyperfunction) solutions as kernel elements.
In the example of ideal MHD, the \emph{resonance singularity} yielded a
current-sheet solution, and its integral defines a singular Casimir element,
by which a \emph{tearing-mode} equilibrium bifurcates
(in the picture of ideal MHD, a tearing mode is stationary because of the flux constraint).  

The mock field was lifted into a physical field by a Hamiltonian that includes it,
while it is initially a mathematical artifact introduced to describe the
Poisson manifold in the higher-dimensional space. Interpreting a Casimir element as an \emph{adiabatic invariant} associated with a hidden ``microscopic'' angle variable, we extended  the phase space by adding the angle variable to the original noncanonical system.   Then, the constancy of the Casimir element was no longer due to the ``topological defect'' (non-trivial kernel) of the Poisson operator, but   due to the symmetry of the Hamiltonian (the newly added angle variable is, of course, ignorable  in the Hamiltonian). 
We  then  unfroze the Casimir element and allowed it to be dynamic by perturbing the Hamiltonian with
a term dependent on the added angle variable.

% Our purpose of delineating the topological constraints on a degenerate Poisson manifold is
% to formulate a singular perturbation that frees the constrains
% ---on a larger system, the sate vector may find regions of lower potential energy and excite instabilities to approach there.

As an explicit application of the formalism, consider the following picture of  the tearing-mode instability in a plasma. 
A tearing mode can be formulated as an equilibrium point on a helical-flux Casimir leaf (see \cite{YD2012}).  
As long as the helical-flux is constrained, the tearing-mode cannot grow.  
Upon  introducing  a perturbation to change the helical flux, as well as to ``dissipate'' the energy, an   
 unstable tearing mode can, then, be formulated as a negative-energy perturbation that can grow by diminishing the energy.   In this picture, the negative energy is absorbed by an ``external system'' through the pathway introduced by the new angle variable;  the extended phase space of the canonized  Hamiltonian system includes this ``external system'' so that the total energy  remains  conserved.  
 
 We envision that many dissipation driven instabilities in fluids and plasma systems can be cast into this basic geometric picture.

%%%%%%%%%%%%%%%%%%%%%%%%%%%%%%%%%%%%%%%%%%%%%%%%%%%%%%%%%%%%%%%%%%%%%%%%%%%%%%%%%%%%%%%%%%%%%%%%%%
%%%%%%%%%%%%%%%%%%%%%%%%%%%%%%%%%%%%%%%%%%%%%%%%%%%%%%%%%%%%%%%%%%%%%%%%%%%%%%%%%%%%%%%%%%%%%%%%%%
%%%%%%%%%%%%%%%%%%%%%%%%%%%%%%%%%%%%%%%%%%%%%%%%%%%%%%%%%%%%%%%%%%%%%%%%%%%%%%%%%%%%%%%%%%%%%%%%%%
%%%%%%%%%%%%%%%%%%%%%%%%%%%%%%%%%%%%%%%%%%%%%%%%%%%%%%%%%%%%%%%%%%%%%%%%%%%%%%%%%%%%%%%%%%%%%%%%%%
%%%%%%%%%%%%%%%%%%%%%%%%%%%%%%%%%%%%%%%%%%%%%%%%%%%%%%%%%%%%%%%%%%%%%%%%%%%%%%%%%%%%%%%%%%%%%%%%%%
% \acknowledgments
\ack
The authors acknowledge discussions with Professor Robert L.\  Dewar and and Professor Yasuhide Fukumoto, and thank them for their suggestions.
The work of ZY was supported by Grant-in-Aid for Scientific Research from
the Japanese Ministry of Education, Science and Culture No.~19340170, while that of PJM was supported by US DOE contract  DE-FG03-96ER-54346.
%%%%%%%%%%%%%%%%%%%%%%%%%%%%%%%%%%%%%%%%%%%%%%%%%%%%%%%%%%%%%%%%%%%%%%%%%%%%%%%%%%%%%%%%%%%%%%%%%%
%%%%%%%%%%%%%%%%%%%%%%%%%%%%%%%%%%%%%%%%%%%%%%%%%%%%%%%%%%%%%%%%%%%%%%%%%%%%%%%%%%%%%%%%%%%%%%%%%%
%%%%%%%%%%%%%%%%%%%%%%%%%%%%%%%%%%%%%%%%%%%%%%%%%%%%%%%%%%%%%%%%%%%%%%%%%%%%%%%%%%%%%%%%%%%%%%%%%%
%%%%%%%%%%%%%%%%%%%%%%%%%%%%%%%%%%%%%%%%%%%%%%%%%%%%%%%%%%%%%%%%%%%%%%%%%%%%%%%%%%%%%%%%%%%%%%%%%%
%%%%%%%%%%%%%%%%%%%%%%%%%%%%%%%%%%%%%%%%%%%%%%%%%%%%%%%%%%%%%%%%%%%%%%%%%%%%%%%%%%%%%%%%%%%%%%%%%%
%             APPENDIX
\appendix
\section{Harmonic field and cohomology}
\label{appendix:cohomology}

We denote by $L^2(\Omega)$ the Hilbert space of Lebesgue-measurable, square-integrable
real vector functions on $\Omega$, which is endowed with the standard inner product
$\langle \bi{a}, \bi{b}\rangle = \int_\Omega \rmd^3x\, \bi{a}\cdot\bi{b}$ and   norm
$\| \bi{a} \| = \langle \bi{a},\bi{a}\rangle^{1/2}$.
We   use the same notation for the $L^2$-norm and inner product,  regardless of the dimensions of
independent and dependent variables, and we also use the standard notation for  Sobolev spaces.

To separate the fixed degrees of freedom pertinent to the fluxes, 
we invoke the Hodge--Kodaira decomposition, with the definitions
\numparts
\begin{eqnarray}
L^2_\sigma(\Omega) = \{ \bi{u}\in L^2(\Omega);\, \nabla\cdot\bi{u}=0,\,
\bi{n}\cdot\bi{u}=0 \},
\label{L2sigma}\\
L^2_\Sigma(\Omega) = \{ \bi{u}\in L^2(\Omega);\, \nabla\cdot\bi{u}=0,\,
\bi{n}\cdot\bi{u}=0,\, \Phi_\ell(\bi{u})=0~(\forall\ell) \}.
\label{L2Sigma}\\
L^2_{\rm H}(\Omega) = \{ \bi{u}\in L^2(\Omega);\, \nabla\times\bi{u}=0,\,\nabla\cdot\bi{u}=0,\,
\bi{n}\cdot\bi{u}=0 \}.
\label{L2H}
\end{eqnarray}
\endnumparts
% The space $L^2_\sigma(\Omega)$ is the totality of magnetic fields that satisfy the boundary condition (\ref{BC}).
The dimension of $L^2_{\rm H}(\Omega)$, the space of \emph{harmonic fields} (or De Rham cohomologies),
is equal to the genus ${m}$ of $\Omega$
and $L^2_{\rm H}(\Omega)$ is spanned by gradients of 
\emph{angle variables} $\theta_\ell$ ($\ell=1,\cdots,{m}$) such that
\begin{equation}
\nabla \theta_\ell \in L^2_{\rm H}(\Omega),
\quad \lshad \theta_\ell\rshad_{\Sigma_\ell} = \theta_\ell |_{\Sigma_\ell^+}-\theta_\ell |_{\Sigma_\ell^-} = 1,
\label{angle}
\end{equation}
where $\Sigma_\ell^\pm$ denote both sides of $\Sigma_\ell$.
For calculational convenience we have normalized the angle by $2\pi$.

We can now state the orthogonal \emph{Hodge-Kodaira decomposition}:
\begin{equation}
L^2_\sigma(\Omega) = L^2_\Sigma(\Omega) \oplus L^2_{\rm H}(\Omega) .
\label{Hodge-Kodaira}
\end{equation}
If $\bi{B}\in L^2_\sigma(\Omega)$ is a magnetic filed,
it can be decomposed into the fixed ``vacuum'' field $\bi{B}_{\rm H}\in L^2_{\rm H}(\Omega)$ (which carries the given fluxes $\Phi_1,\cdots,\Phi_{m}$) 
and a residual component $\bi{B}_\Sigma\in L^2_\Sigma(\Omega)$ driven by currents within the volume $\Omega$.
% \begin{equation}
% \bi{B} = \bi{B}_\Sigma + \bi{B}_{\rm H},
% \quad [\bi{B}_\Sigma={\cal P}_\Sigma \bi{B}\in L^2_\Sigma(\Omega),\, \bi{B}_{\rm H}\in L^2_{\rm H}(\Omega)],
% \label{decomposition-B}
% \end{equation}
% where ${\cal P}_\Sigma$ denotes the orthogonal projector from $L^2(\Omega)$ onto $L^2_\Sigma(\Omega)$.

The components $\bi{B}_\Sigma$ and $\bi{B}_{\rm H}$ can be represented uniquely, up to arbitrary constants, respectively,  by a vector potential $\bi{A}_\Sigma \in L^2_\Sigma(\Omega)$ and a (multi-valued) scalar potential $\sum_{\ell}j_{\ell}\theta_{\ell}$, where the ``periods'' $j_{\ell}$ give the loop integrals of $\bi{B}_{\rm H}$ through the handles of $\Omega$; 
so that, by Amp\`ere's law, the periods are proportional to currents external to $\Omega$. 
% (However, in the following development we instead represent $\bi{B}_{\rm H}$ in terms of a vector potential $\bi{A}_{\rm H} \in L^2_\Sigma(\Omega)$ --- see Lemma~\ref{lemma:T-operator}.) 

% We can decompose the harmonic field $\bi{B}_{\rm H}$ from $\bi{B}$ and consider the
% remaining $\bi{B}_\Sigma$ to be a dynamical variable; the flux conservation law 
% restricts $\bi{B}_{\rm H}$ to be constant.
% Hence, a coadjoint orbit is contained in $L_\Sigma(\Omega)$.
% Naturally, we thus expect that the fluxes $\Phi_\ell(\bi{B})$ ($\ell=1,\cdots,{m}$) are Casimir elements.

%%%%%%%%%%%%%%%%%%%%%%%%%%%%%%%%%%%%%%%%%%%%%%%%%%%%%%%%%%%%%%%%%%%%%%%%%%%%%%%%%%%%%%%%%%%%%%%%%%
%%%%%%%%%%%%%%%%%%%%%%%%%%%%%%%%%%%%%%%%%%%%%%%%%%%%%%%%%%%%%%%%%%%%%%%%%%%%%%%%%%%%%%%%%%%%%%%%%%
%%%%%%%%%%%%%%%%%%%%%%%%%%%%%%%%%%%%%%%%%%%%%%%%%%%%%%%%%%%%%%%%%%%%%%%%%%%%%%%%%%%%%%%%%%%%%%%%%%
%%%%%%%%%%%%%%%%%%%%%%%%%%%%%%%%%%%%%%%%%%%%%%%%%%%%%%%%%%%%%%%%%%%%%%%%%%%%%%%%%%%%%%%%%%%%%%%%%%
%%%%%%%%%%%%%%%%%%%%%%%%%%%%%%%%%%%%%%%%%%%%%%%%%%%%%%%%%%%%%%%%%%%%%%%%%%%%%%%%%%%%%%%%%%%%%%%%%%
%   REFERENCES
%%%%%%%%%%%%%%%%%%%%%%%%%%%%%%%%%%%%%%%%%%%%%%%%%%%%%%%%%%%%%%%%%%%%%%%%%%%%%%%%%%%%%%%%%%%%%%%%%%
\section*{References}
%\begin{references}

\end{document}